\documentclass[a4paper,12pt]{article}

\usepackage{amsmath}
\usepackage{amssymb}
\usepackage{color}
\usepackage{graphics}
\usepackage{graphicx,epsfig}
\usepackage{amsfonts}

\usepackage{inputenc}
\inputencoding{latin1}

\newcommand{\bsg}{BR(b\rightarrow s \gamma)}
\newcommand{\DM}{\Omega_{CDM}h^2}

\newcommand{\gmu}{\delta a_{\mu}}

\newcommand{\stau}{\tilde{\tau}}

\newcommand{\neut}{\tilde{\chi}^0_1}
\newcommand{\nneut}{\tilde{\chi}^0_2}
\newcommand{\nnneut}{\tilde{\chi}^0_3}
\newcommand{\nnnneut}{\tilde{\chi}^0_4}

\newcommand{\charg}{\tilde{\chi}^+_1}
\newcommand{\ccharg}{\tilde{\chi}^+_2}
\newcommand{\DeltaO}{\Delta^{\Omega}}

\newcommand{\tanb}{\tan\beta}
\newcommand{\sigI}{1\sigma}
\newcommand{\sigII}{2\sigma}

\def\r2{\sqrt 2}
\def\beq{\begin{equation}}
\def\eeq{\end{equation}}
\def\beqn{\begin{eqnarray}}
\def\eeqn{\end{eqnarray}}
\def\sinW2{\sin^2\theta_W}

\def\mz2{M_{z}^2}
\def\c2b{\cos 2\beta}

\def\mz{M_Z}

\begin{document}
  
  \begin{titlepage}
    \vspace*{-15mm}
    \begin{flushright}
      \today\\
    \end{flushright}
    \vspace*{5mm}

    \begin{center}
      {
	\sffamily \Large Natural Dark Matter in SUSY GUTs
	with Non-universal Gaugino Masses
      }
      \\[8mm]
      S.~F.~King$^{a,}$\footnote{E-mail: \texttt{sfk@hep.phys.soton.ac.uk}},
      J.~P.~Roberts$^{b,}$\footnote{E-mail: \texttt{roberts@fuw.edu.pl}}
      and D.~P.~Roy$^{c,d,}$\footnote{E-mail: \texttt{dproy1@gmail.com}}
      \\[3mm]
      
	{\small $^a$ School of Physics and Astronomy, University of
	  Southampton\\ $^b$ Institute of Theoretical Physics,
	  Warsaw University,\\ 00-681 Warsaw, Poland\\ $^c$ Homi
	  Bhabha Centre for Science Education,\\ Tata Institute of
	  Fundamental Research,\\ Mumbai 400088,
	  India\\ $^d$ Instituto de Fisica Corpuscular, CSIC-U.de
	  Valencia, Correos,\\ E-46071 Valencia, Spain } \\
	\end{center}
	
    \vspace*{0.75cm}
    
    \begin{abstract}
      \noindent
      We consider neutralino dark matter within the framework of SUSY
      GUTs with non-universal gaugino masses. In particular we focus
      on the case of $SU(5)$ with a SUSY breaking F-term in the 1, 24,
      75 and 200 dimensional representations.  We discuss the 24 case
      in some detail, and show that the bulk dark matter region cannot
      be accessed.  We then go on to consider the admixture of the
      singlet SUSY breaking F-term with one of the 24, 75 or 200
      dimensional F-terms, and show that in these cases it becomes
      possible to access the bulk regions corresponding to low
      fine-tuned dark matter.  Our results are presented in the
      $(M_1,M_2)$ plane for fixed $M_3$ and so are useful for
      considering general GUT models, as well as more general
      non-universal gaugino models.
    \end{abstract}
  \end{titlepage}

  \newpage

  \setcounter{footnote}{0}
  \renewcommand{\theequation}{\arabic{section}.\arabic{equation}} 
 
  \section{Introduction}

  Supersymmetry (SUSY) at the TeV scale remains an attractive
  possibility for new physics beyond the Standard Model. SUSY helps in
  the unification of couplings in Grand Unified Theories (GUTs), and
  provides a resolution of some aspects of the hierarchy problem. In
  addition the lightest SUSY particle (LSP) may be a neutralino
  consisting of a linear combination of Bino, Wino and neutral
  Higgsinos, providing a consistent WIMP dark matter candidate
  \cite{hep-ph/0312378}.  For example the minimal supersymmetric
  standard model (MSSM) with conserved R-parity provides such an LSP
  with a mass of order the electroweak scale. Although general
  arguments suggest that such a particle should provide a good dark
  matter candidate \cite{hep-ph/9506380}, the successful regions of
  parameter space allowed by WMAP and collider constraints are now
  tightly restricted \cite{hep-ph/0411216}-\cite{hep-ph/0402240}.

  Such a restricted parameter space has lead to recent claims that
  supersymmetry must be fine-tuned to fit the observed dark matter
  relic density \cite{hep-ph/0601041}. This is a serious concern for
  supersymmetry, especially as much of the motivation for
  supersymmetry arises from fine-tuning arguments in the form of its
  solution to the hierarchy problem. In previous work
  \cite{hep-ph/0603095}-\cite{hep-ph/0609147} we quantitatively
  studied the fine-tuning cost of the primary dark matter regions
  within the MSSM. It was found that the majority of dark matter
  regions did indeed require some degree of fine-tuning, and that this
  fine-tuning could be directly related to the mechanism responsible
  for the annihilation of SUSY matter in the early universe that
  defined each region. The one region that exhibited no fine-tuning at
  all was the `bulk region' in which the dominant annihilation
  mechanism is via t-channel slepton exchange. This region can be
  accessed in models in which the gauginos have non-universal soft
  masses at the GUT scale \cite{hep-ph/0407218}.

  These results motivate a more careful study of models that give rise
  to non-universal gaugino masses. In our previous work such a region
  was accessed by allowing all the gaugino masses to vary
  independently. Such an approach is very unconstrained. We would
  expect the gaugino masses to arise from a deeper theory such as
  string constructions, as studied in \cite{hep-ph/0608135},
  \cite{hep-ph/0609147} or in GUT
  models~\cite{etcEllis:1985jn}-\cite{Huitu:1999vx}. Both approaches
  generally impose specific relations between the gaugino masses at
  the GUT scale. In this paper we shall discuss non-universal gaugino
  masses in a more general way than previously, allowing for different
  relative signs of gaugino masses, focusing on $SU(5)$ GUTs as an
  example, although it is clear that similar effects can be achieved
  in other GUTs such as $SO(10)$ or Pati-Salam.  We shall show how the
  bulk region may be readily accessed in such models providing that
  the SUSY breaking sector arises from a combination of an $SU(5)$
  singlet 1, together with an admixture of one of the 24, 75 or 200
  representations of $SU(5)$. We will also show that in all cases the
  fine-tuning required to access such a region remains small.

  The rest of the paper is set out as follows.  First we review our
  methodology in section \ref{sec:methods}. In section \ref{sec:SU(5)}
  we review the structure of gaugino non-universality in $SU(5)$. In
  section \ref{sec:24} we consider the specific case where all of the
  gaugino masses arise from a 24 of $SU(5)$. In section
  \ref{sec:admixtures} we generalise this to the case where the masses
  arise from an admixture of the singlet representation and one of the
  24, 75 or 200. In section \ref{sec:Conc} we present our conclusions.

  \section{Methodology}
  \label{sec:methods}

  \subsection{Codes}

  The GUT structure of the theory is a structure that is imposed on
  the soft SUSY breaking masses at the GUT scale, $m_{GUT}\approx
  2\times 10^{16}$~GeV. To study the low energy phenomenology of such
  a model we need to run the mass spectrum down to the electroweak
  scale. To do this we use the RGE code {\tt SoftSusy}
  \cite{hep-ph/0104145}. This interfaces with the MSSM package within
  {\tt micrOMEGAs} \cite{hep-ph/0112278}. We use this to calculate the
  dark matter relic density $\DM$, as well as $\bsg$ and $\gmu$.

  \subsection{Experimental Bounds}

  Not all choices of parameters are equal. After running the mass
  spectrum of the model point from the GUT scale to the electroweak
  scale we perform a number of checks. A point is ruled out if it: 

  \begin{enumerate}
  \item doesn't provide radiative electroweak symmetry breaking
    (REWSB).
  \item violates mass bounds on particles from the Tevatron and LEP2.
  \item results in a lightest supersymmetric particle (LSP) that is
    not the lightest neutralino.
  \end{enumerate}

  In the remaining parameter space we plot regions that fit $\bsg$ and
  $\gmu$ at $\sigI$ and $\sigII$.

  \subsubsection{$\gmu$}

  Present measurements of the value of the anomalous magnetic moment
  of the muon $a_\mu$ deviate from the theoretical calculation of the
  SM value\footnote{There is a long running debate as to whether the
  calculation of the hadronic vacuum polarisation in the Standard Model
  should be done with the $e^+e^-$ data, or the $\tau$. The weight of
  evidence indicates the $e^+e^-$ data is more reliable and we use
  this in our work.}. Taking the current experimental world average,
  and state of the art Standard Model value from \cite{hep-ph/0703049}
  there is a discrepancy:
  \begin{equation}
    (a_\mu)_{exp}-(a_\mu)_{SM}=\delta a_\mu = (2.95\pm
    0.88)\times 10^{-9}
  \end{equation}
  which amounts to a 3.4$\sigma$ deviation from the Standard Model
  value.

  \subsubsection{$\bsg$}

  The variation of $\bsg$ from the value predicted by the Standard
  Model is highly sensitive to SUSY contributions arising from charged
  Higgs-top loops and chargino-stop loops. To date no deviation from
  the Standard Model has been detected. We take the current world
  average from \cite{hfag} of the BELLE \cite{hep-ex/0103042}, CLEO
  \cite{hep-ex/0108033} and BaBar \cite{Aubert:2005cu} experiments:
  \begin{equation}
    \bsg = (3.55 \pm 0.26) \times 10^{-4}
  \end{equation}

  \subsubsection{$\DM$}
  Evidence from the CMB and rotation curves of galaxies both point to
  a large amount of cold non-baryonic dark matter in the universe. The
  present measurements \cite{astro-ph/0603449}
  place the dark matter density at:
  \begin{equation}
    \DM = 0.106 \pm 0.008
  \end{equation}
  
  For any point that lies within the $\sigII$ allowed region we
  calculate the fine-tuning and plot the resulting colour-coded point.

  \subsection{Fine-tuning}
  As in \cite{hep-ph/0603095} we follow Ellis and Olive
  \cite{Ellis:2001zk} in quantifying the fine-tuning price of fitting
  dark matter with the measure:
  \begin{equation}
    \Delta_a^{\Omega}=\left|\frac{\partial
      \ln\left(\DM\right)}{\partial \ln\left(a \right)}\right|
    \label{meas}
  \end{equation}
  where we take the total fine-tuning of a point to be equal to the
  largest individual tuning, $\Delta=\text{max}(\Delta_a)$.

  \section{Gaugino Non-universality in $SU(5)$}
  \label{sec:SU(5)}

  In the non-universal $SU(5)$ model \cite{hep-ph/0304108}, in
  addition to the singlet F-term SUSY breaking, the gauge kinetic
  function can also depend on a non-singlet chiral superfield $\Phi$,
  whose auxiliary $F$-component acquires a large vacuum expectation
  value (vev). In general the gaugino masses come from the following
  dimension five term in the Lagrangian:
  \begin{equation}
    L={\frac{{<F_\Phi>}_{ij}}{M_{Planck}}} \lambda_i \lambda_j
  \end{equation}
  where $\lambda_{1,2,3}$ are the $U(1)$, $SU(2)$ and $SU(3)$ gaugino
  fields i.e. the bino $\tilde B$, the wino $\tilde W$ and the gluino
  $\tilde g$ respectively.  Since the gauginos belong to the adjoint
  representation of $SU(5)$, $\Phi$ and $F_\Phi$ can belong to any of
  the irreducible representations appearing in their symmetric
  product, i.e.
  \begin{equation}
    {(24 \times 24)}_{symm} =1+24+75+200
  \end{equation}
  The minimal supergravity (mSUGRA) model assumes $\Phi$ to be a
  singlet, which implies equal gaugino masses at the GUT scale. On the
  other hand if $\Phi$ belongs to one of the non-singlet
  representations of $SU(5)$, then these gaugino masses are unequal
  but related to one another via the representation invariants.  Thus
  the three gaugino masses at the GUT scale in a given representation
  $n$ are determined in terms of a single SUSY breaking mass parameter
  $m_{1/2}$ by
  \begin{eqnarray}
    M_{1,2,3} = C^n_{1,2,3} m_{1/2}
    \label{relativegauginos}
  \end{eqnarray}
  where $C^{1}_{1,2,3}=(1,1,1)$, $C^{24}_{1,2,3}=(-1,-3,2)$, $C^{75}_{1,2,3}=(-5,3,1)$ and
  $C^{200}_{1,2,3}=(10,2,1)$.  The resulting ratios of
  $M_i$'s for each $n$ are listed
  in Table~\ref{tabrelativeweights}.
  \begin{table}[ht]
    {\centering
      \begin{tabular}[h]{|c| c c c|}
	\hline
	$n$ & $M_3$ & $M_2$ & $M_1$ \\
	\hline
	\hline
	1 & 1 & 1 & 1 \\
	24 & 1 & $-3/2$ & $-1/2$ \\
	75 & 1 & 3 & $-5$  \\
	200 & 1 & 2 & 10  \\
	\hline
      \end{tabular}
      \par}
    \centering
    \caption{Relative values of the SU(3), SU(2) and U(1)
      gaugino masses at GUT scale for different
      representations $n$ of the chiral superfield $\Phi$.}
    \label{tabrelativeweights}
  \end{table}
  Of course in general the gauge kinetic function can involve several
  chiral superfields belonging to different representations of $SU(5)$
  which gives us the freedom to vary mass ratios continuously. In
  this, more general, case we can parameterise the GUT scale gaugino
  masses as:
  \begin{equation}
    M_{1,2,3} = C^n_{1,2,3} m^n_{1/2}
  \end{equation}
  where $m^n_{1/2}$ is the soft gaugino mass arising from the $F$-term
  vev in the representation $n$.

  These non-universal gaugino mass models are known to be consistent
  with the observed universality of the gauge couplings at the GUT
  scale~\cite{etcEllis:1985jn}-\cite{Huitu:1999vx},~\cite{Chattopadhyay:2001mj}
  \begin{equation}
  \alpha_3=\alpha_2=\alpha_1=\alpha (\simeq 1/25)
  \end{equation}
  Since the gaugino masses evolve like the gauge couplings at one loop
  level of the renormalisation group equations (RGE), the three
  gaugino masses at the electroweak scale are proportional to the
  corresponding gauge couplings, i.e.
  \begin{eqnarray}
  M_1^{EW} & = & (\alpha_1/\alpha_G) M_1  \simeq  (25/60) C_1^n m^n_{1/2} \nonumber \\
  M_2^{EW} & = & (\alpha_2/\alpha_G) M_2 \simeq  (25/30)C_2^n m^n_{1/2} \nonumber \\
  M_3^{EW} & = & (\alpha_3/\alpha_G) M_3  \simeq  (25/9) C_3^n m^n_{1/2}
  \label{gauginoEW}
  \end{eqnarray}

  For simplicity we shall assume a universal SUSY breaking scalar mass
  $m_0$ at the GUT scale.  Then the corresponding scalar masses at the
  EW scale are given by the renormalisation group evolution
  formulae~\cite{Carena:1994bv}.

  \section{The 24 model}
  \label{sec:24}

  We have previously seen \cite{hep-ph/0603095} that a ratio
  $M_1:M_2:M_3=0.5:1:1$ allows us to access the bulk region without
  violating LEP bounds. The bulk region in the CMSSM is usually ruled
  out because of a light Higgs. By allowing $M_3$ to be large we can
  avoid a light Higgs while allowing $M_1$ to be light enough to give
  a light bino neutralino and light sleptons. This enhances neutralino
  decay via light t-channel slepton exchange and gives access to the
  bulk region.

  From Table~\ref{tabrelativeweights} we observe that only the
  24 model predicts a mass ratio $M_1<M_3$. Therefore we shall explore
  the 24 model first.  For the 24 model we have the input parameters:
  \begin{equation*}
    a \in \left\{m_0,\ m^{24}_{1/2},\ A_0,\ \tan\beta,\
    \text{sign}(\mu) \right\}.
  \end{equation*}
  where the masses are all set as in the CMSSM except for the gaugino
  masses which have the form:
  \begin{eqnarray*}
    M_1&=&-0.5 ~m^{24}_{1/2}\\
    M_2&=&-1.5 ~m^{24}_{1/2}\\
    M_3&=&m^{24}_{1/2}
  \end{eqnarray*}

  With this gaugino mass structure, the bino mass in the 24 for a
  given $m_{1/2}$ is half of the bino mass in the CMSSM for the same
  $m_{1/2}$. The bino mass also affects the running of the slepton
  masses such that lower $M_1$ corresponds to a lower slepton
  mass. Therefore the 24 will have lower mass sleptons than the CMSSM
  for a given value of $m_0$ and $m_{1/2}$. Light sleptons enhance the
  annihilation of neutralinos via t-channel slepton exchange (giving
  rise to a WMAP region known as the bulk region). Therefore we expect
  the bulk region to appear at larger $m_{1/2}$ than in the CMSSM and
  thus circumvent the Higgs mass bound.

  \begin{figure}[ht]
    \begin{minipage}{0.8\textwidth}
      \begin{minipage}{0.4\textwidth}
	\begin{center}
	\scalebox{0.65}{\includegraphics{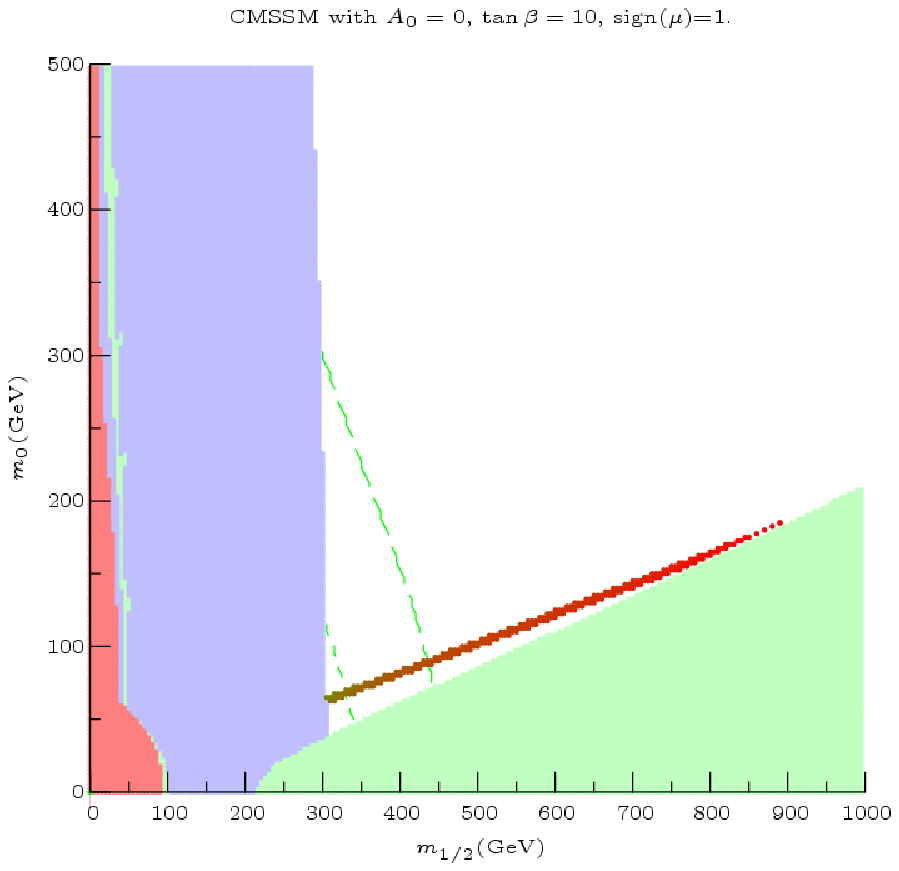}}
	\end{center}
      \end{minipage}
      \hfill
      \begin{minipage}{0.4\textwidth}
	\begin{center}
	\scalebox{0.65}{\includegraphics{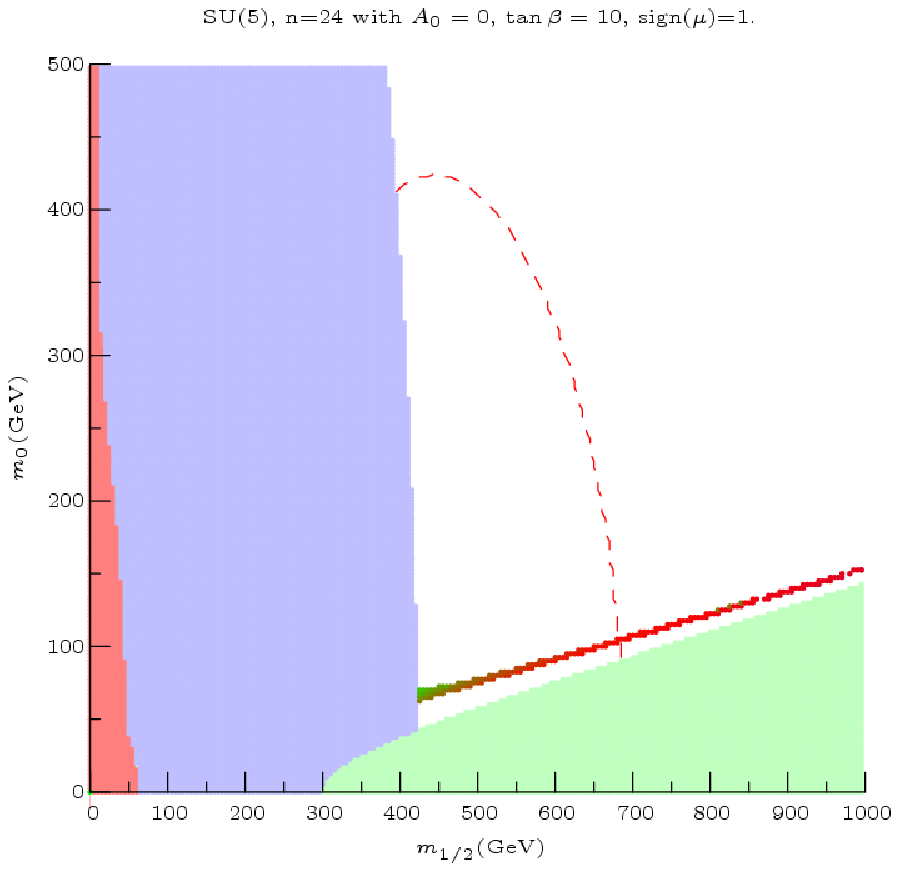}}
	\end{center}
      \end{minipage}
      
      \vspace*{0.5cm}
      \begin{minipage}{0.4\textwidth}
	\begin{center}
	\scalebox{0.65}{\includegraphics{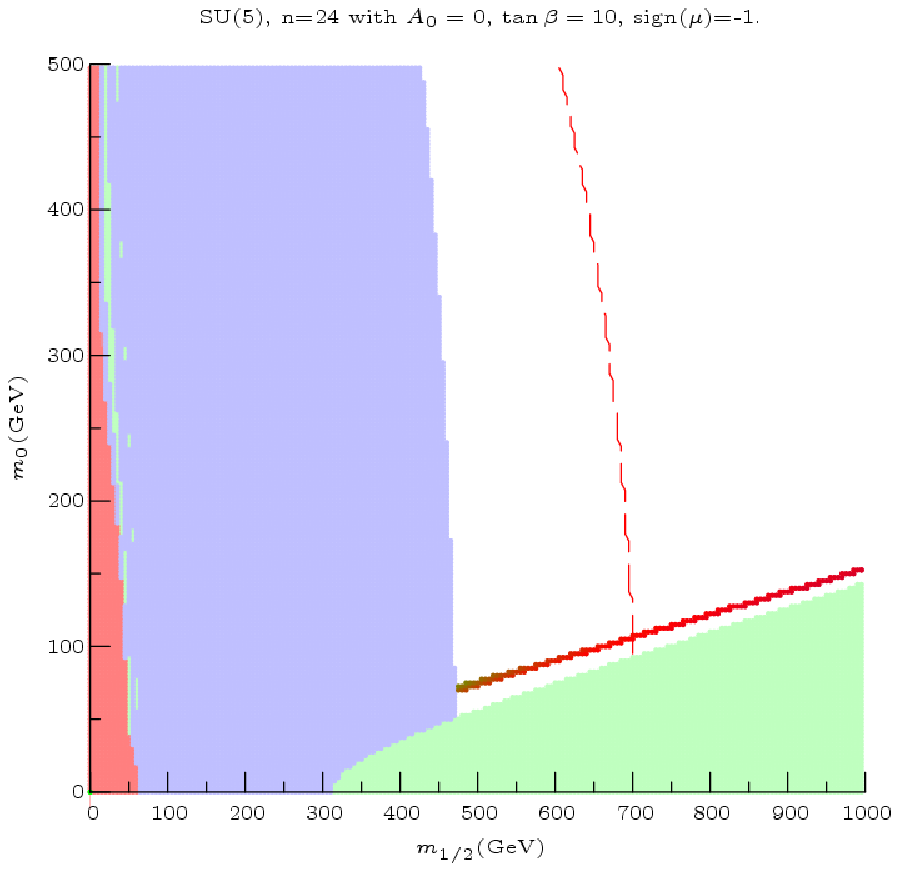}}
	\end{center}
      \end{minipage}
      \hfill
      \begin{minipage}{0.4\textwidth}
	\begin{center}
	  \scalebox{0.65}{\includegraphics{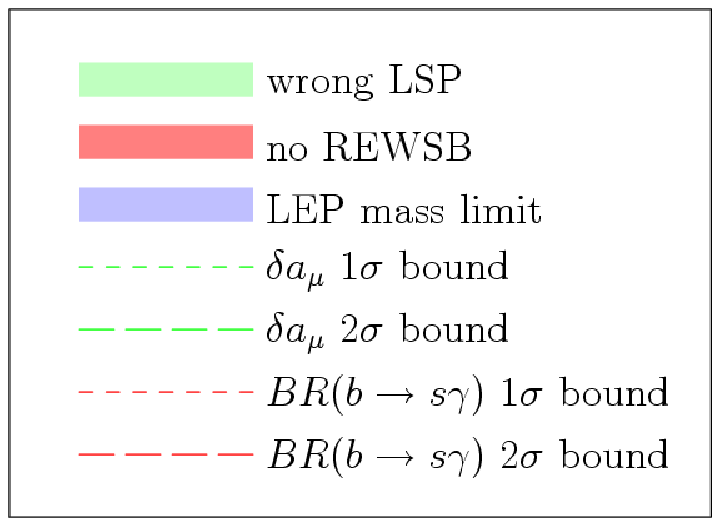}}
	\end{center}
      \end{minipage}
    \end{minipage}
    \hspace*{0.5cm}
    \hfill
    \begin{minipage}{0.1\textwidth}
      \scalebox{0.8}{\includegraphics{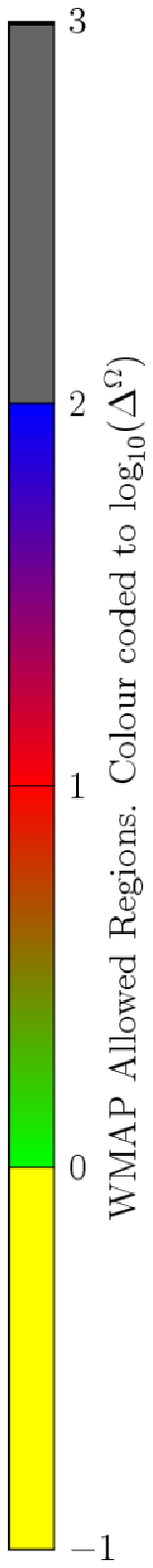}}
    \end{minipage}

    \caption{\small The parameter space for the CMSSM (top-left), the 24
      model with sign$(\mu)$ $+$ve (top-right) and with sign$(\mu)$
      $-$ve (bottom). Low $m_0$ is ruled out as the $\stau$ becomes the
      LSP(light green). Low $m_{1/2}$ is ruled out as $m_h<111$GeV. In
      the remaining parameter space, the only strip of allowed dark
      matter is a $\stau-\neut$ coannihilation strip which shows
      comparable degrees of tuning in all plots. \label{A0,0,tanb,10}}
  \end{figure}
  
  To study this effect, we look at the $(m_0,~m_{1/2})$ plane in both
  the CMSSM and the 24 in Fig.~\ref{A0,0,tanb,10}. The CMSSM is shown
  in the top-left panel, the $24$ with $\mu$ positive in the top-right
  panel and the $24$ with $\mu$ negative is shown in the bottom-left
  panel.

  In the CMSSM scan we can see that low $m_0$ is ruled out as the stau
  becomes lighter than the neutralino. Low $m_{1/2}$ is ruled out as
  $m_h<111$GeV. The contours of $1$ and $2\sigma$ for $\gmu$ (green
  short and long dashed lines respectively) are plotted in the
  remaining parameter space, showing that the current measurement of
  $\gmu$ favours low $m_0$ and $m_{1/2}$. Finally the region that
  satisfies WMAP is plotted as a multicoloured strip that runs
  alongside the light green region ruled out by a stau LSP. This WMAP
  strip is mostly red. This colour coding refers to a log measure of
  the fine-tuning and can be read off via the log-scale on the right
  hand side. The tuning of the $\stau$ coannihilation strip agrees
  with our previous findings.

  In the second and third panels of Fig.~\ref{A0,0,tanb,10} we once
  again display the $(m_0,~m_{1/2})$ plane but this time using the 24
  model's soft gaugino masses with $\mu$ positive and negative
  respectively. In both cases, low $m_0$ is ruled out by a stau LSP
  and low $m_{1/2}$ is ruled out by a light Higgs. 

  The $\gmu$ and $\bsg$ values are significantly different in the 24
  model than in the CMSSM. Firstly neither 24 plot has a region that
  agrees with the current measured value of $\gmu$ (they both give
  $\gmu \pm\mathcal{O}(10^{-10})$). Secondly $\bsg$ becomes an important
  constraint. For $\mu$ $+$ve, the model agrees with the measured
  value of $\bsg$ at $1\sigma$ for large $m_{1/2}(> 700$~GeV) and
  agrees at $2\sigma$ for low $m_{1/2}$. With $\mu$ $-$ve, only the
  parameter space at $m_0>700$~GeV fits $\bsg$ at $\sigII$. Lower
  $m_0$ exceeds this limit.

  Now consider the change in the dark matter strip. We expected to be
  able to access the bulk region in this model as we would have a
  lighter bino neutralino and lighter sleptons in the 24 model than in
  the CMSSM. This should move the bulk region to larger values of
  $m_{1/2}$ and out from under the region ruled out by the LEP2 bound
  on the lightest Higgs boson. 

  Contrary to our naive expectations, though the bulk region has moved
  to larger $m_{1/2}$ in the 24 model, it remains ruled out. This is
  because the gaugino mass relations in the 24 also result in a
  lighter Higgs mass than the CMSSM, for the same $m_0,~m_{1/2}$.  The
  only difference between the CMSSM and the 24 model is the magnitude
  and sign of the $M_1$ and $M_2$ gaugino masses. Therefore the Higgs
  mass must be sensitive either to the sign difference between
  $M_{1,2}$ and $M_3$ or the larger value of $M_2$.

  First consider the effect of the relative sign between $M_{1,2}$ and
  $M_3$. In most RGEs the gaugino masses appear squared, however the
  trilinear RGEs have the form:
  \begin{equation}
    \frac{dA_t}{dt}=\frac{1}{8\pi^2}\left[6|Y_t|^2A_t+ |Y_b|^2 A_b+
      \left(\frac{16}{3}g_3^2M_3+ 3g_2^2M_2+\frac{13}{15}g_1^2M_1\right)\right]
  \end{equation}

  \begin{figure}[ht]
    \begin{center}
      \scalebox{0.75}{\includegraphics{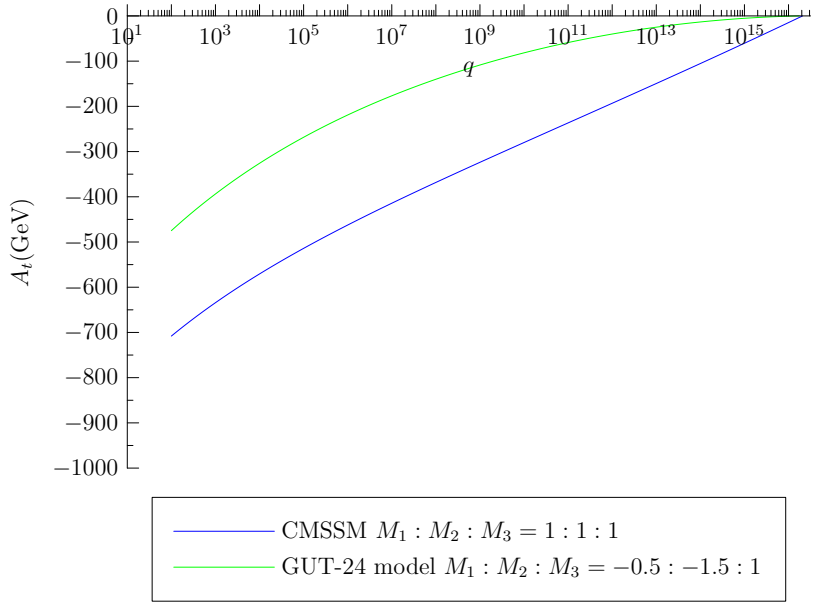}}
    \end{center}
    \caption{\small Here we show the running of $A_t$ from the GUT scale value
      of $A_t=0$ to the weak scale for the point $m_0=100$GeV,
      $m_{1/2}=350$GeV, $\tanb=10$, $A_0=0$. The running for the CMSSM
      is shown in blue, the running for the 24 model is shown in
      red. \label{A0,0,tanb,10,AtRun}}
  \end{figure}

  If all $M_i$ are positive, then the gauginos provide a large
  positive contribution to the RGE and so help to push the trilinear
  negative through the running. This in turn affects the running of
  the Higgs mass. In the 24 case, the sign of $M_{1,2}$ are opposite
  to that of $M_3$ and so they reduce the contribution from the
  Gauginos and thus reduce the magnitude of the running, resulting in
  a small trilinear at the electroweak scale. Now we note that the
  contribution of $M_{1,2}$ are suppressed relative to that of $M_3$
  by a factor of $g_i^2$, but this is partially compensated by the
  fact that $|M_2|>|M_3|$ at the GUT scale. Therefore both the sign
  and magnitude of $M_2(\text{GUT})$ are responsible for a substantial
  change in the running of the trilinears. This is shown in
  Fig.~\ref{A0,0,tanb,10,AtRun}.

  The change in the trilinear affects the running of $m_{H_u}^2$ via
  the RGE:
  \begin{eqnarray}
    \nonumber \frac{dm_{H_u}^2}{dt}&=&\frac{1}{8\pi^2}\bigg[
      3\left|Y_t\right|^2\left(m_{Q_3}^2
      +m_{U_3}^2+m_{H_u}^2+\left|A_t\right|^2\right)\\
      &&\left.-\left(3g_2^2\left|M_2\right|^2
      +\frac{3}{5}g_1^2\left|M_1\right|^2\right)\right]
  \end{eqnarray}

  A smaller top trilinear results in a smaller running of the Higgs
  mass and a lighter Higgs. Therefore, as the 24 model results in a
  smaller value of $A_t$ at all energies below the GUT scale, it gives
  a smaller mass for the lightest Higgs than for the same model point
  in the CMSSM. This means that the LEP mass bounds for the lightest
  Higgs are more restrictive in the 24 model than in the
  CMSSM. Unfortunately, this results in the LEP Higgs bound ruling
  out the bulk region for all interesting regions of parameter space
  of the 24 model.

  \section{Two $SU(5)$ Sectors}
  \label{sec:admixtures}

  We have seen that neither the CMSSM, corresponding to a singlet SUSY
  breaking sector, nor the 24 model is capable of accessing the bulk
  region of neutralino parameter space. Equally, as the 75 and 200
  models have $|M_1|>|M_3|$, these sectors are even worse. In this section
  we therefore consider the next simplest possibility, namely that of
  two different SUSY breaking $SU(5)$ representations acting together.
  Indeed, once one has accepted the existence of a single 24, 75 or
  200 dimensional SUSY breaking sector, it seems perfectly natural to
  allow the standard singlet SUSY breaking sector at the same time. In
  practice it may be difficult to avoid this scenario.

  Therefore we shall focus on the three simplest scenarios. We take
  the cases of a SUSY breaking sector consisting of:
  \begin{center}
    \begin{tabular}{ll}
      A & $(1+24)$\\
      B & $(1+75)$\\
      C & $(1+200)$
    \end{tabular}
  \end{center}

  If we were to extend our model to allow three or four $SU(5)$
  representations contributing to SUSY breaking at once, we would be
  able to produce any pattern of non-universal gaugino masses. By
  constraining our model to two sectors we provide restrictions on the
  choice of gaugino masses which makes access to the bulk region
  non-trivial, and provides insight into what ingredients are required
  to achieve it.

  \begin{table}[ht]
    \begin{center}
      \begin{tabular}{|l|l|l|l|}
	\hline Mass & A ($1+24$) & B ($1+75$) & C $(1+200)$\\ 
	\hline
	$M_1$ & $m^1_{1/2}-0.5~m^{24}_{1/2}$ & $m^1_{1/2}-5~m^{75}_{1/2}$
	& $m^1_{1/2}+10~m^{200}_{1/2}$ \\ 
	$M_2$ & $m^1_{1/2}-1.5~m^{24}_{1/2}$ & $m^1_{1/2}+3~m^{75}_{1/2}$
	& $m^1_{1/2}+2~m^{200}_{1/2}$\\ 
	$M_3$ & $m^1_{1/2}+m^{24}_{1/2}$ & $m^1_{1/2}+m^{75}_{1/2}$
	& $m^1_{1/2}+m^{200}_{1/2}$\\ 
	\hline
      \end{tabular}
    \end{center}
    \caption{The gaugino mass relations for the different $(1+n)$ SUSY
      breaking scenarios.\label{tab:gauginoMasses}}
  \end{table}

  Within these models, we have different gaugino mass relations, shown
  in Table.~\ref{tab:gauginoMasses}. By varying the soft gaugino masses
  $m^{1,n}_{1/2}$, we describe three planes in the $M_{1,2,3}$
  parameter space.

  \begin{figure}[!ht]
    \begin{center}
      \scalebox{0.75}{\includegraphics{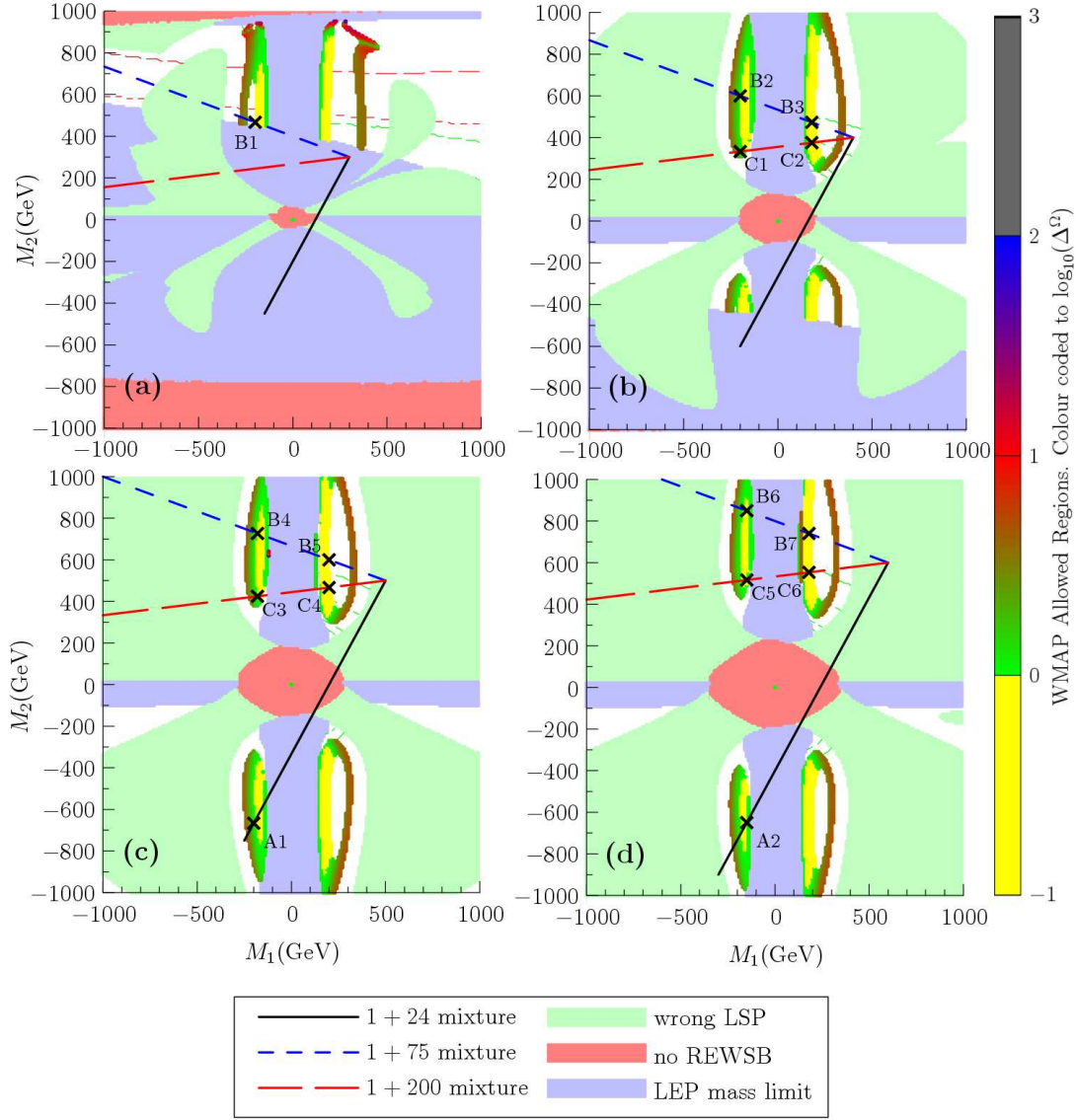}}
    \end{center}
    \caption{\small The $(M_1,~M_2)$ plane with non-universal gaugino
    masses defined at the GUT scale. We take $m_0=70$~GeV, $A_0=0$ and
    $\tanb=10$ throughout vary $M_3$: (a) $M_3=300$~GeV, (b)
    $M_3=400$~GeV, (c) $M_3=500$~GeV, (d) $M_3=600$~GeV. For fixed
    $M_3$, the allowed parameter space for each GUT mixture plotted as
    a line the $(M_1,~M_2)$ parameter space. The WMAP allowed regions
    correspond to the elliptical regions in each quadrant, and are
    partially obscured by disallowed regions in panels (a) and
    (b). The $\bsg$ and $\gmu$ regions are displayed as in
    Fig.~\ref{A0,0,tanb,10} and discussed in the
    text.
    \label{fig:admixtures}}
  \end{figure}

  Our aim is to access the bulk region. In \cite{hep-ph/0603095} we
  found that the bulk region can be accessed in a model with
  non-universal gaugino masses for $m_0=50-80$~GeV. Therefore we fix
  $m_0=70$~GeV, $A_0=0$ and $\tanb=10$. In
  Figs.~\ref{fig:admixtures}(a)-(d) we plot the $(M_1,~M_2)$ plane for
  increasing values of $M_3$, from $300-600$~GeV. As $M_1$ and $M_2$
  can in general be either positive or negative in $(1+n)$ scenarios,
  we allow $M_1$ and $M_2$ to take positive and negative values. For a
  given $M_3$, the gaugino mass relation of
  Table~\ref{tab:gauginoMasses} constrain each of the $(1+n)$
  scenarios to a line in the $(M_1,~M_2)$ plane. We plot these lines
  for each case.

  As each model has the singlet representation as a limit when
  $m^n_{1/2}\rightarrow 0$, all the lines converge at a point. At this
  point the model is precisely that of the CMSSM, and as such is ruled
  out for almost all $M_3$ by a $\stau$ LSP or the LEP bound on the
  lightest Higgs. The other end of each line corresponds to the
  opposite limit $m^1_{1/2}=0,~m^n_{1/2}=M_3$.

  We also plot the $\bsg$ and $\gmu$ constraints. The only region that
  doesn't fit $\bsg$ within $\sigII$ is panel (a) at large $M_2$. The
  values of $\gmu$ are insensitive to $M_3$. In the quadrant with
  $M_1$ and $M_2$ +ve we have the largest SUSY contribution to $\gmu$,
  enabling the model to fit $\gmu$ at $\sigI$. In the quadrant with
  $M_1$ +ve, $M_2$ -ve, the model can fit $\gmu$ at $\sigII$. For
  negative $M_1$ we get a negative SUSY contribution, $\gmu$. If we
  were to plot the parameter space with $\mu$ negative, $\gmu$ would
  have the opposite sign and the model would fit the observed value of
  $\gmu$ for negative $M_1$.

  Finally, we plot the dark matter regions with colours corresponding
  to their fine-tuning calculated with respect to the general
  non-universal gaugino model with parameters:
  $a\in\left\{m_0,~M_1,~M_2,~M_3,~A_0,~\tanb\right\}$. This allows us
  to easily pick out the bulk region as it is `supernatural' with
  $\DeltaO<1$ and is therefore plotted in yellow. We use this to pick
  out the points at which each $(1+n)$ representation provides access
  to the bulk region. We take these points and calculate the dark
  matter fine-tuning with respect to the $(1+n)$ model in question.

  \begin{table}[ht]
    \begin{center}
      \begin{tabular}{|l|l|l|l|l|}
	\hline
	\multicolumn{1}{|c}{Parameter}&
	\multicolumn{1}{|c}{A1}&
	\multicolumn{1}{c|}{}&
	\multicolumn{1}{|c}{A2}&
	\multicolumn{1}{c|}{}\\
	\cline{2-5}
	& value & $\DeltaO$ & value & $\DeltaO$ \\
	\hline
	$m_0$ & 70 & 1.43 & 70 & 0.96 \\
	$m^1_{1/2}$ & 33.3 & 0.026 & 100 & 0.39 \\
	$m^{24}_{1/2}$ & 466.7 & 0.075 & 500 & 1.02 \\
	$A_0$ & 0 & 0 & 0 & 0 \\
	$\tanb$ & 10 & 0.37 & 10 & 0.21\\
	\hline
	Max & & 1.43 & & 0.96\\
	\hline\hline
	$M_1$ & -200 & 0.19 & -150 & 0.59\\
	$M_2$ & -666.7 & 0.21 & -650 & 0.38 \\
	$M_3$ & 500 & 0.075 & 600 & 0.0088 \\
	\hline
      \end{tabular}
    \end{center}
    \caption{\small The fine-tuning for points A1 and A2 that lie
    within the bulk region for the $(1+24)$ model. For both points
    $m^{24}_{1/2}>m^1_{1/2}$, so the gaugino masses arise
    predominantly from the $24$. In the lower section of the table we
    give the corresponding GUT scale $M_i$ for each point. As the
    tunings plotted in Fig.~\ref{fig:admixtures} are calculated with
    respect to the parameter set
    $a\in\left\{m_0,~M_1,~M_2,~M_3,~A_0,~\tanb\right\}$, we give the
    relevant tunings with respect to the individual $M_i$ for
    comparison.\label{tab:1+25}}
  \end{table}

  First consider the $1+24$ model. In
  Figs.~\ref{fig:admixtures}(a),~(b) the model does not access the
  bulk region. This fits with our results of section \ref{sec:24} as
  low $m_{1/2}$ is ruled out by a light Higgs in the 24 scenario. In
  Figs.~\ref{fig:admixtures}(c),~(d), we can access the bulk region
  with a mixture that is primarily 24. We show the corresponding
  fine-tuning for both points in Table~\ref{tab:1+25}. Note that for
  both points $m^{24}_{1/2}>m^{1}_{1/2}$, so the gaugino masses arise
  predominantly from the $24$.

  Next consider the $1+75$ model. This model lies along the blue short
  dashed line. The $75$ limit is not shown. This is because in the
  pure $75$ scenario $M_1=-5M_3$. Therefore the $75$ limit lies
  outside the range plotted for all $M_3$ that we consider. In such
  a limit, as studied in \cite{hep-ph/0210034},~\cite{hep-ph/0304108},
  the lightest neutralino is predominantly higgsino. As discussed
  earlier we cannot access the bulk region in such a limit. This limit
  lies off the plots and we do not consider it further here.

  \begin{table}[ht]
      \begin{tabular}{|l|l|l|l|l|l|l|l|l|}
	\hline
	\multicolumn{1}{|c}{Parameter}&
	\multicolumn{1}{|c}{B1}&
	\multicolumn{1}{c|}{}&
	\multicolumn{1}{|c}{B2}&
	\multicolumn{1}{c|}{}&
	\multicolumn{1}{|c}{B3}&
	\multicolumn{1}{c|}{}&
	\multicolumn{1}{|c}{B4}&
	\multicolumn{1}{c|}{}\\
	\cline{2-9}
	 & value & $\DeltaO$ & value & $\DeltaO$ & value & $\DeltaO$ & value & $\DeltaO$ \\
	\hline
	$m_0$ & 70 & 0.91 & 70 & 1.18 & 70 & 0.86 & 70 & 1.0  \\
	$m^1_{1/2}$ & 217 & 0.78 & 300 & 0.64 & 363 & 1.4 & 387 & 1.1  \\
	$m^{75}_{1/2}$ & 83.3 & 1.4 & 100 & 0.91 & 36.7 & 0.67 & 113 & 1.5 \\
	$A_0$ & 0 & 0 & 0 & 0 & 0 & 0 & 0 & 0  \\
	$\tanb$ & 10 & 0.13 & 10 & 0.29 & 10 & 0.14 & 10 & 0.32  \\
	\hline
	Max & & 1.4 & & 0.91 & & 1.4 & & 1.5 \\
	\hline\hline
	$M_1$ & -200 & 0.66 & -200 & 0.38 & 180 & 0.67 & -180 & 0.51 \\
	$M_2$ & 467 & 0.086 & 600 & 0.032 & 473 & 0.096 & 727 & 0.075  \\
	$M_3$ & 300 & 0.13 & 400 & 0.071 & 400 & 0.061 & 500 & 0.047  \\
	\hline
      \end{tabular}
      \begin{tabular}{|l|l|l|l|l|l|l|}
	\hline
	\multicolumn{1}{|c}{Parameter}&
	\multicolumn{1}{|c}{B5}&
	\multicolumn{1}{c|}{}&
	\multicolumn{1}{|c}{B6}&
	\multicolumn{1}{c|}{}&
	\multicolumn{1}{|c}{B7}&
	\multicolumn{1}{c|}{}\\
	\cline{2-7}
	 & value & $\DeltaO$ & value & $\DeltaO$ & value & $\DeltaO$ \\
	\hline
	$m_0$  & 70 & 0.75 & 70 & 0.95 & 70 & 0.84 \\
	$m^1_{1/2}$  & 450 & 1.8 & 475 & 1.7 & 530 & 2.0\\
	$m^{75}_{1/2}$  & 50 & 0.99 & 125 & 2.4 & 70 & 1.2 \\
	$A_0$  & 0 & 0 & 0 & 0 & 0 & 0 \\
	$\tanb$  & 10 & 0.15 & 10 & 0.32 & 10 & 0.22 \\
	\hline
	Max  & & 1.8 & & 2.4 & & 2.0 \\
	\hline\hline
	$M_1$  & 200 & 0.80 & -150 & 0.55 & 180 & 0.64 \\
	$M_2$  & 600 & 0.038 & 850 & 0.082 & 740 & 0.031 \\
	$M_3$  & 500 & 0.014 & 600 & 0.16 & 600 & 0.12 \\
	\hline
      \end{tabular}
    \caption{\small The fine-tuning for points B1-7 that lie within the bulk
    region for the $(1+75)$ model. For all points
    $m^{75}_{1/2}<m^1_{1/2}$, so the gaugino masses arise predominantly
    from the singlet.  In the lower section of the table we
    give the corresponding GUT scale $M_i$ for each point. As the
    tunings plotted in Fig.~\ref{fig:admixtures} are calculated with
    respect to the parameter set
    $a\in\left\{m_0,~M_1,~M_2,~M_3,~A_0,~\tanb\right\}$, we give the
    relevant tunings with respect to the individual $M_i$ for
    comparison.\label{tab:1+75}}
  \end{table}

  In the $75$, $M_1$ is negative. This results in two scenarios in
  which $M_1<M_3$. For a small $m^{75}_{1/2}$, the negative
  contribution results in a small, positive, $M_1$. For a slightly
  larger $m^{75}_{1/2}$, we get a small, negative $M_1$. This is shown
  in the plots and is the reason that the $1+75$ accesses the bulk
  region twice for most values of $M_3$, once for each sign of
  $M_1$. We study the 7 resulting points in the bulk regions in
  Table~\ref{tab:1+75}. Note that for all points
  $m^{75}_{1/2}<m^{1}_{1/2}$, so the gaugino masses arise
  predominantly from the singlet.

  Finally consider the case of the $1+200$ model. The lines
  corresponding to this model are plotted in red with long dashes. As
  in the $1+75$ case, in the $200$ limit the lightest neutralino is
  higgsino and we cannot access the bulk region. This limit lies off
  the plots and we do not consider it further here.

  \begin{table}[ht]
    \begin{center}
      \begin{tabular}{|l|l|l|l|l|l|l|}
	\hline
	\multicolumn{1}{|c}{Parameter}&
	\multicolumn{1}{|c}{C1}&
	\multicolumn{1}{c|}{}&
	\multicolumn{1}{|c}{C2}&
	\multicolumn{1}{c|}{}&
	\multicolumn{1}{|c}{C3}&
	\multicolumn{1}{c|}{}\\
	\cline{2-7}
	  & value & $\DeltaO$ & value & $\DeltaO$ & value & $\DeltaO$ \\
	\hline
	$m_0$  & 70 & 1.6 & 70 & 0.89 & 70 & 1.1  \\
	$m^1_{1/2}$  & 467 & 0.11 & 424 & 1.7 & 576 & 1.4 \\
	$m^{200}_{1/2}$  & -66.7 & 0.40 & -24.4 & 0.93 & -75.6 & 2.2  \\
	$A_0$  & 0 & 0 & 0 & 0 & 0 & 0  \\
	$\tanb$  & 10 & 0.79 & 10 & 0.25 & 10 & 0.54  \\
	\hline
	Max  & & 1.6 & & 1.7 & & 2.2 \\
	\hline\hline
	$M_1$  & -200 & 0.19 & 180 & 0.67 & -180 & 0.56 \\
	$M_2$  & 333 & 0.83 & 376 & 0.31 & 424 & 0.59  \\
	$M_3$  & 400 & 0.75 & 400 & 0.22 & 500 & 0.39  \\
	\hline
	\hline
	\multicolumn{1}{|c}{Parameter}&
	\multicolumn{1}{|c}{C4}&
	\multicolumn{1}{c|}{}&
	\multicolumn{1}{|c}{C5}&
	\multicolumn{1}{c|}{}&
	\multicolumn{1}{|c}{C6}&
	\multicolumn{1}{c|}{}\\
	\cline{2-7}
	  & value & $\DeltaO$ & value & $\DeltaO$ & value & $\DeltaO$ \\
	\hline
	$m_0$   & 70 & 0.78 & 70 & 0.97 & 70 & 0.86 \\
	$m^1_{1/2}$   & 533 & 2.3 & 683 & 2.3 & 647 & 2.5\\
	$m^{200}_{1/2}$   & -33.3 & 1.3 & -83.3 & 3.1 & -46.7 & 1.7 \\
	$A_0$   & 0 & 0 & 0 & 0 & 0 & 0 \\
	$\tanb$   & 10 & 0.23 & 10 & 0.43 & 10 & 0.25 \\
	\hline
	Max   & & 2.3 & & 3.1 & & 2.5 \\
	\hline\hline
	$M_1$   & 200 & 0.80 & -150 & 0.59 & 180 & 0.63 \\
	$M_2$   & 467 & 0.25 & 517 & 0.49 & 553 & 0.22 \\
	$M_3$   & 500 & 0.13 & 600 & 0.20 & 600 & 0.047 \\
	\hline
      \end{tabular}
    \end{center}
    \caption{\small The fine-tuning for points C1-6 that lie within
    the bulk region for the $(1+200)$ model. For all points
    $|m^{200}_{1/2}|<|m^1_{1/2}|$, so the gaugino masses arise
    predominantly from the $1$. We also give the corresponding GUT
    scale $M_i$ for each point. As the tunings in
    Fig.~\ref{fig:admixtures} are calculated with respect to the
    parameters $a\in\left\{m_0,~M_1,~M_2,~M_3,~A_0,~\tanb\right\}$, we
    give the tunings with respect to $M_i$ for
    comparison.\label{tab:1+200}}
  \end{table}

  As the $200$ has all gaugino masses positive, and large $M_1$, we
  cannot access the bulk region in the $200$ limit. However by
  combining with the singlet we can get $|M_1|<|M_3|$ by taking a
  small, negative $m^{200}_{1/2}$. This allows such a model to access
  the bulk region for positive and negative small $M_1$. We study the
  resulting 6 points in the bulk region in Table~\ref{tab:1+200}.In
  all points $|m^{200}_{1/2}|<|m^1_{1/2}|$ so the gaugino masses arise
  predominantly from the $1$.

  \begin{table}
    \begin{center}
      \begin{tabular}{|l|l|}
	\hline 
	Particle & Mass (GeV)\\ 
	\hline 
	$\neut$ (bino)               & 78.1 \\
	$\nneut$ (wino)              & 457  \\ 
	$\nnneut$ (higgsino)         & 614  \\ 
	$\nnnneut$ (higgsino)        & 636  \\ 
	$\charg$ (wino)              & 461  \\ 
	$\ccharg$ (higgsino)         & 635  \\ 
	\hline
	$M^{EW}_1$                   & 81   \\
	$M^{EW}_2$                   & 470  \\
	$M^{EW}_3$                   & 1120 \\
	$\mu$                        & 611  \\ 
	\hline
	$\tilde{g}$                  & 1150 \\
	$\tilde{\tau}_1$             & 104  \\
	$\tilde{\tau}_2$             & 399  \\
	$\tilde{e}_R,\tilde{\mu}_R$  & 115  \\
	$\tilde{e}_L,\tilde{\mu}_L$  & 399  \\
	$\tilde{t}_1$                & 793  \\
	$\tilde{t}_2$                & 1025 \\
	$\tilde{b}_1$                & 980  \\
	$\tilde{b}_2$                & 1000  \\
	$\tilde{q}_{1,2,R}$          & $\sim 1005$ \\
	$\tilde{q}_{1,2,L}$          & $\sim 1070$ \\
	\hline
      \end{tabular}
    \end{center}
    \caption{\small The SUSY mass spectrum of point B5 from
    Fig.~\ref{fig:admixtures}. This spectrum is characteristic of all
    bulk region points we have studied. We display the hierarchy and
    flavour of the neutralino and chargino sectors. We also display
    the values of the neutralino mass parameters for completeness. For
    the squarks we take a typical squark mass rather than list the
    full squark spectrum. The exceptions are the 3rd family squarks
    that we list separately. Finally, the sneutrinos are degenerate
    with $\tilde{e},\tilde{\mu}_L$. \label{tab:spectrum}}
  \end{table}

  The hierarchy of the weak scale SUSY spectrum is fairly stable for
  all the points shown in
  Fig~\ref{fig:admixtures}. Table~\ref{tab:spectrum} lists the
  neutralino, chargino and sfermion masses along with $M_1$, $M_2$ and
  the Higgsino mass parameter $\mu$ for the point B5 as an example. In
  contrast to the CMSSM the bino is lighter than the wino by a factor
  of 6. Correspondingly the right and left slepton masses are split by
  a large factor.  The small value of $m_0$ also ensures that the
  right handed sleptons are considerably lighter than the wino. Hence
  a large fraction of wino decay is predicted to proceed via
  $\stau_1$, resulting in one or more tau leptons in the final state
  in addition to the missing-$E_T$. Though the light selectron and
  smuon have negligible left-handed components, and so cannot take
  part in the wino decay, the heavier selectron and smuon are still
  lighter than the wino in all points we consider. A wino decay via a
  left-handed selectron/smuon would give a distinctive signal in the
  form of hard electron(s)/muon(s) in addition to the
  missing-$E_T$. Thus one expects a distinctive SUSY signal from
  squark/gluino cascade decays at LHC containing hard isolated leptons
  in addition to the missing-$E_T$ and jets.

  \section{Conclusions}
  \label{sec:Conc}
  In previous work we found that a model with non-universal gaugino
  masses could access the bulk region in which t-channel slepton
  exchange alone could account for the observed dark matter relic
  density. The bulk region is an attractive prospect as it allows SUSY
  to account for the observed dark matter relic density without any
  appreciable fine-tuning. However, a model with entirely free gaugino
  masses is very unconstrained. Such non-universality must arise from
  a deeper structure and such structures should impose restrictions on
  the precise form of the gaugino masses at the GUT scale.
  
  In this paper we have considered neutralino dark matter within the framework
  of SUSY GUTs with non-universal gaugino masses. 
  We have taken the specific case of an $SU(5)$ GUT model where the
  gaugino masses arise from different irreducible representations of
  the symmetric product of the adjoint representations. 
  In particular we 
  focused on the case of $SU(5)$ with a SUSY breaking F-term in
  the 1, 24, 75 and 200 dimensional representations.  We discussed
  the 24 case in some detail, and showed that the bulk dark matter
  region cannot be accessed in this case. In general if we just
  take the simplest case in which the gaugino masses arise from only
  one representation, we find that as far as achieving the
  bulk region is concerned, there is no advantage over the
  CMSSM. This is in part due to the surprising result that the sign
  and magnitude of $M_2$ with respect to $M_3$ has an important effect
  on the lightest Higgs mass through its effect on the top trilinear.

  We then went on to consider the case of the singlet SUSY breaking
  F-term combined with an admixture of one of the 24, 75 or 200
  dimensional F-terms. Such a scenario is natural once we allow the
  higher dimensional representations in our theory.  In all these
  cases we showed that it becomes possible to access the bulk regions
  corresponding to low fine-tuned dark matter. In addition, the degree
  of fine-tuning required to access the bulk region remains small in
  the GUT models. Therefore we conclude that such models can access
  the bulk region and naturally account for the observed dark matter
  relic density. 

  Finally we note that the results in Fig.~\ref{fig:admixtures} are
  presented in the $(M_1,M_2)$ plane for fixed $M_3$ and so are useful
  for considering general GUT models, as well as more general
  non-universal gaugino models. The hierarchy of weak scale SUSY
  spectrum is fairly stable for all the points shown in
  Fig.~\ref{fig:admixtures}. Both the right and left sleptons are
  lighter than the wino, implying a large leptonic BR of wino
  decay. This promises a distinctive SUSY signal from squark/gluino
  cascade decays at LHC in the form of hard isolated leptons in
  addition to the missing-$E_T$ and jets.

  \section*{Acknowledgements}
  SFK would like to thank the Warsaw group for its hospitality and
  support under the contract MTKD-CT-2005-029466. The work of JPR was
  funded under the FP6 Marie Curie contract MTKD-CT-2005-029466. The
  work of DPR is partly supported by MEC grants FPA2005-01269,
  SAB2005-0131.


\begin{thebibliography}{36}

  \bibitem{hep-ph/0312378}
    D.~J.~H.~Chung, L.~L.~Everett, G.~L.~Kane, S.~F.~King, J.~D.~Lykken and L.~T.~Wang,
    Phys.\ Rept.\ {\bf 407} (2005) 1
    [arXiv:hep-ph/0312378].

  \bibitem{hep-ph/9506380}
    G.~Jungman, M.~Kamionkowski and K.~Griest,
    Phys.\ Rept.\ {\bf 267} (1996) 195
    [arXiv:hep-ph/9506380].

  \bibitem{hep-ph/0411216}
    J.~R.~Ellis, S.~Heinemeyer, K.~A.~Olive and G.~Weiglein,
    JHEP {\bf 0502} (2005) 013
    [arXiv:hep-ph/0411216].

  \bibitem{hep-ph/0407218}
    G.~Belanger, F.~Boudjema, A.~Cottrant, A.~Pukhov and A.~Semenov,
    Nucl.\ Phys.\ B {\bf 706} (2005) 411
    [arXiv:hep-ph/0407218].

  \bibitem{hep-ph/0403214}
    H.~Baer, A.~Belyaev, T.~Krupovnickas and A.~Mustafayev,
    JHEP {\bf 0406} (2004) 044
    [arXiv:hep-ph/0403214].

  \bibitem{hep-ph/0307389}
    M.~R.~Ramage and G.~G.~Ross,
    JHEP {\bf 0508} (2005) 031
    [arXiv:hep-ph/0307389].

  \bibitem{hep-ph/9407404}
    M.~Olechowski and S.~Pokorski,
    Phys.\ Lett.\ B {\bf 344} (1995) 201
    [arXiv:hep-ph/9407404].

  \bibitem{hep-ph/9412379}
    F.~M.~Borzumati, M.~Olechowski and S.~Pokorski,
    Phys.\ Lett.\ B {\bf 349} (1995) 311
    [arXiv:hep-ph/9412379].

  \bibitem{hep-ph/9508249}
    V.~Berezinsky, A.~Bottino, J.~R.~Ellis, N.~Fornengo, G.~Mignola and S.~Scopel,
    Astropart.\ Phys.\ {\bf 5} (1996) 1
    [arXiv:hep-ph/9508249].

  \bibitem{hep-ph/9701301}
    P.~Nath and R.~Arnowitt,
    Phys.\ Rev.\ D {\bf 56} (1997) 2820
    [arXiv:hep-ph/9701301].

  \bibitem{hep-ph/0001019}
    E.~Accomando, R.~Arnowitt, B.~Dutta and Y.~Santoso,
    Nucl.\ Phys.\ B {\bf 585} (2000) 124
    [arXiv:hep-ph/0001019].

  \bibitem{hep-ph/0210339}
    R.~Arnowitt and B.~Dutta,
    arXiv:hep-ph/0210339.

  \bibitem{hep-ph/0304101}
    R.~Dermisek, S.~Raby, L.~Roszkowski and R.~Ruiz De Austri,
    JHEP {\bf 0304} (2003) 037
    [arXiv:hep-ph/0304101].

  \bibitem{hep-ph/0210034}
    V.~Bertin, E.~Nezri and J.~Orloff,
    JHEP {\bf 0302} (2003) 046
    [arXiv:hep-ph/0210034].

  \bibitem{hep-ph/0211071}
    A.~Birkedal-Hansen and B.~D.~Nelson,
    Phys.\ Rev.\ D {\bf 67} (2003) 095006
    [arXiv:hep-ph/0211071].

  \bibitem{hep-ph/0304108}
    U.~Chattopadhyay and D.~P.~Roy,
    Phys.\ Rev.\ D {\bf 68} (2003) 033010
    [arXiv:hep-ph/0304108].

  \bibitem{hep-ph/0509275}
    D.~G.~Cerdeno, K.~Y.~Choi, K.~Jedamzik, L.~Roszkowski and R.~Ruiz de Austri,
    arXiv:hep-ph/0509275.

  \bibitem{hep-ph/0402240}
    L.~Covi, L.~Roszkowski, R.~Ruiz de Austri and M.~Small,
    JHEP {\bf 0406} (2004) 003
    [arXiv:hep-ph/0402240].

  \bibitem{hep-ph/0601041}
    N.~Arkani-Hamed, A.~Delgado and G.~F.~Giudice,
    arXiv:hep-ph/0601041.

  \bibitem{hep-ph/0603095}
    S.~F.~King and J.~P.~Roberts,
    JHEP {\bf 0609} (2006) 036
    [arXiv:hep-ph/0603095].

  \bibitem{hep-ph/0608135}
    S.~F.~King and J.~P.~Roberts,
    JHEP {\bf 0701} (2007) 024
    [arXiv:hep-ph/0608135].

  \bibitem{hep-ph/0609147}
    S.~F.~King and J.~P.~Roberts,
    Acta Phys.\ Polon.\  B {\bf 38} (2007) 607
    [arXiv:hep-ph/0609147].

  \bibitem{etcEllis:1985jn}
    \label{etcEllis:1985jn}
    J.~R.~Ellis, K.~Enqvist, D.~V.~Nanopoulos and K.~Tamvakis,
    Phys.\ Lett.\ B {\bf 155}, 381 (1985);  M.~Drees,
    Phys.\ Lett.\ B {\bf 158}, 409 (1985).



  \bibitem{etcAnderson:1996bg}
    G.~Anderson, C.~H.~Chen, J.~F.~Gunion, J.~Lykken, T.~Moroi and Y.~Yamada,
    arXiv:hep-ph/9609457;  G.~Anderson, H.~Baer, C.~h.~Chen and X.~Tata,
    Phys.\ Rev.\ D {\bf 61}, 095005 (2000).


  \bibitem{Huitu:1999vx}
    K.~Huitu, Y.~Kawamura, T.~Kobayashi and K.~Puolamaki,
    Phys.\ Rev.\ D {\bf 61}, 035001 (2000).

  \bibitem{hep-ph/0104145}
    B.~C.~Allanach,
    Comput.\ Phys.\ Commun.\ {\bf 143} (2002) 305
    [arXiv:hep-ph/0104145].

  \bibitem{hep-ph/0112278} G.~Belanger, F.~Boudjema, A.~Pukhov and
    A.~Semenov, 
    Comput.\ Phys.\ Commun.\ {\bf 149} (2002)
    103 [arXiv:hep-ph/0112278]; 
    G.~Belanger, F.~Boudjema, A.~Pukhov and A.~Semenov,
    Comput.\ Phys.\ Commun.\ {\bf 176}
    (2007) 367 [arXiv:hep-ph/0607059].  

  \bibitem{hep-ph/0703049}
    J.~P.~Miller, E.~de Rafael and B.~L.~Roberts,
    Rept.\ Prog.\ Phys.\  {\bf 70} (2007) 795
    [arXiv:hep-ph/0703049].


  \bibitem{hfag} Heavy Flavour Averaging Group,
    www.slac.stanford.edu/xorg/hfag.

  \bibitem{hep-ex/0103042}
    K.~Abe {\it et al.} [Belle Collaboration],
    Phys.\ Lett.\ B {\bf 511} (2001) 151
    [arXiv:hep-ex/0103042];
    P.~Koppenburg {\it et al.}  [Belle Collaboration],
    Phys.\ Rev.\ Lett.\  {\bf 93} (2004) 061803
    [arXiv:hep-ex/0403004].

  \bibitem{hep-ex/0108033}
    D.~Cronin-Hennessy {\it et al.} [CLEO Collaboration],
    Phys.\ Rev.\ Lett.\ {\bf 87} (2001) 251808
    [arXiv:hep-ex/0108033].

  \bibitem{Aubert:2005cu}
    B.~Aubert {\it et al.}  [BABAR Collaboration],
    Phys.\ Rev.\  D {\bf 72} (2005) 052004
    [arXiv:hep-ex/0508004].
    B.~Aubert {\it et al.}  [BaBar Collaboration],
    Phys.\ Rev.\ Lett.\  {\bf 97} (2006) 171803
    [arXiv:hep-ex/0607071].

  \bibitem{astro-ph/0603449}
    D.~N.~Spergel {\it et al.}  [WMAP Collaboration],
    arXiv:astro-ph/0603449.


  \bibitem{Ellis:2001zk}
    J.~R.~Ellis and K.~A.~Olive,
    Phys.\ Lett.\ B {\bf 514} (2001) 114
    [arXiv:hep-ph/0105004].


  \bibitem{Chattopadhyay:2001mj}
    \label{Chattopadhyay:2001mj}
    U.~Chattopadhyay and P.~Nath,
    Phys.\ Rev.\ D {\bf 65}, 075009 (2002).


  \bibitem{Carena:1994bv}
    See e.g. M.~Carena, M.~Olechowski, S.~Pokorski and C.~E.~Wagner,
    Nucl.\ Phys.\ B {\bf 426}, 269 (1994).

  \end{thebibliography}
\end{document}